\documentclass[11pt,a4paper]{article}

\usepackage[T1]{fontenc}
\usepackage[utf8]{inputenc}
\usepackage{lmodern}
\usepackage[margin=1in]{geometry}
\usepackage{amsmath,mathtools}
\usepackage{amsfonts}
\usepackage{amsthm}
\usepackage{graphicx}
\usepackage{booktabs}
\usepackage{xcolor}
\usepackage{xspace}
\usepackage{tikz}
\usepackage{subcaption}
\usepackage{cite}
\usepackage{xurl}
\usepackage[hidelinks]{hyperref}
\usepackage{orcidlink}
\usepackage[capitalize,nameinlink,noabbrev]{cleveref}
\usetikzlibrary{patterns}

\newtheorem{theorem}{Theorem}
\newtheorem{lemma}[theorem]{Lemma}

\newtheorem{observation}[theorem]{Observation}

\newcommand{\US}{\ensuremath{\mathsf{US}}}

\newcommand{\clusters}{\ensuremath{\mathcal{C}}}
\newcommand{\upbound}{\ensuremath{d^{3/2}}}

\title{Sparse Matrix Multiplication Exponent 1.5 in the Low-bandwidth Model}

\author{
    Chetan Gupta\,\orcidlink{0000-0002-0727-160X}\\
    IIT Roorkee, Roorkee, India\\
    \texttt{chetan.gupta@cs.iitr.ac.in}
    \and
    Jukka Suomela\,\orcidlink{0000-0001-6117-8089}\\
    Aalto University, Helsinki, Finland\\
    \texttt{jukka.suomela@aalto.fi}
    \and
    Hossein Vahidi\,\orcidlink{0000-0002-0040-1213}\\
    Aalto University, Helsinki, Finland\\
    \texttt{hossein.vahidi@aalto.fi}
}
\date{}
\hypersetup{pdftitle={Sparse Matrix Multiplication Exponent 1.5 in the Low-bandwidth Model},pdfauthor={Chetan Gupta, Jukka Suomela, Hossein Vahidi}}

\begin{document}

\maketitle

\begin{abstract}
	We study sparse matrix multiplication in the low-bandwidth model: 
	$n$ computers multiply two $n\times n$ matrices, and each computer can send and receive one $O(\log n)$-bit message per round.
	The input matrices are uniformly sparse, meaning that there are at most $d$ nonzero entries in each row and column, and the task is to compute a uniformly sparse part of their product.
		
	In prior work, Gupta et al.~(SPAA 2022) presented an $O(d^{1.907})$-round algorithm over fields for this problem when the sparsity structure is globally known in advance (this is referred to as the supported setting). The upper bound was later improved to $O(d^{1.832})$ rounds.
	
	In this work, we give a deterministic algorithm that solves this problem in $O(d^{3/2})$ rounds in the supported setting, improving the previous bounds over fields and semirings.


\end{abstract}

\section{Introduction} \label{sec:intro}

In the AI era, \emph{matrix multiplication} is arguably one of the most important computational primitives that we need to be able to solve efficiently in massive scale. Even modest improvements in the efficiency of practical matrix multiplication could potentially yield global energy savings that are measured not in watts or kilowatts but in power plants. Furthermore, modern LLM architectures have many opportunities for exploiting \emph{sparsity}, e.g.\ in the sparse attention layer.

In this work we study the complexity of sparse matrix multiplication with massively parallel algorithms. We have multiple ``machines'' (an abstraction that can represent e.g.\ threads, warps, cores, processors, or computers as appropriate) that collaborate in solving the task. We assume that \emph{communication} is the key bottleneck; here communication refers to the exchange of information between the machines (and this can represent anything ranging from shared memory accesses to sending messages over a network between computing nodes). The key question we aim to answer is this:
\begin{quote}
How many communication steps are sufficient for the machines to collectively calculate $X = AB$?
\end{quote}

In the theory of distributed computing community, the theoretical framework that has traditionally been used for exploring such questions has been the \emph{congested clique} model \cite{censor-hillel-kaski-etal-2019-algebraic-methods-in-the}. The standard setting is as follows: we have $n$ machines; the task is to calculate $X = AB$ for matrices with $n \times n$ elements; initially each machine holds one row of $A$ and one row of $B$; and in the end each machine has to hold one row of $X$. In each communication round, each computer can send and receive $O(n)$ words (we will assume here that a word is large enough to hold a matrix element or index). It is known that for \emph{dense} matrices we can solve the task in $O(n^{1-2/\omega+\epsilon})$ rounds, where $\omega$ is the familiar matrix multiplication exponent; by plugging in the latest news $\omega \le 9/4$ \cite{openai-2026-an-upper-bound-of-9-4-for-the-matrix,amanov-van-den-berg-zuiddam-2026-multiplying-matrices-via-polynomials} we obtain $O(n^{1/9+\epsilon})$ rounds.

However, when we start to study how much \emph{sparsity} helps, this model is a dead end. Even moderately sparse matrices can be multiplied in this model in $O(1)$ rounds, for boring reasons, as the number of words communicated per round is so large. In particular, the model is useless for comparing different sparse matrix multiplication algorithms, as it will only tell us that both of the candidate algorithms run in $O(1)$ rounds.

A much more useful abstraction for studying how much sparsity helps is a low-bandwidth version of the same model (models of that spirit have been called e.g.\ node-capacitated clique or node-congested clique in the literature). The precise formalism used in recent work \cite{gupta-hirvonen-etal-2022-sparse-matrix-multiplication,gupta-korhonen-etal-2025-low-bandwidth-matrix} is this:
\begin{itemize}
  \item We have $n$ machines, and the task is to calculate $X = AB$ for $n \times n$ matrices.
  \item The matrices $A$ and $B$ are uniformly sparse: there are at most $d$ nonzero elements per row and per column.
  \item We only care about a uniformly sparse part of the product $X$: there are at most $d$ elements per each row and column that we need to calculate.
  \item Initially each machine holds one row of each input matrix, and in the end each machine has to hold one row of the result.
  \item In each communication round, each computer can send and receive only $O(1)$ words.
  \item The sparsity structure is known in advance (supported model), but the values of the matrix elements are only known at runtime.
\end{itemize}
In this formalism, we have conveniently a single parameter $d$ that captures how sparse a product we are computing. In this setting we can solve the task in $O(d^2)$ rounds with a trivial brute-force algorithm, and $\Omega(d)$ is the trivial information-theoretic lower bound. The key question is how much better we can do?

Over fields, prior work managed to push it first to $O(d^{1.907})$ rounds \cite{gupta-hirvonen-etal-2022-sparse-matrix-multiplication} and then to $O(d^{1.832})$ rounds \cite{gupta-korhonen-etal-2025-low-bandwidth-matrix}. We could slightly improve the exponent further by using the latest value of $\omega$, which gives a $O(d^{1.82223})$-round algorithm, but it will become even less practical then. There are fundamental barriers that prevent major advances following the general strategy from \cite{gupta-hirvonen-etal-2022-sparse-matrix-multiplication,gupta-korhonen-etal-2025-low-bandwidth-matrix}.

But today we have fortunately access to the problem-solving and algorithm-design skills of modern LLMs. In this work we present an AI-discovered simple and elegant algorithm that solves the problem in \textbf{\boldmath $O(d^{1.5})$ rounds}.

Notably, the algorithm does not use any machinery from fast centralized matrix multiplication algorithms, its complexity does not depend on $\omega$, it solves matrix multiplication over arbitrary semirings, and it is arguably \textbf{practical}.

In this work we do our best of explaining the AI-discovered algorithm to our fellow humans. While the algorithm design was done by AI (OpenAI GPT-6 primarily), this work is written by humans, for humans.

\paragraph{Open questions for future work.}

Similar to prior work \cite{gupta-hirvonen-etal-2022-sparse-matrix-multiplication,gupta-korhonen-etal-2025-low-bandwidth-matrix}, our new algorithm assumes that the sparsity structure is known in advance, i.e., we have a \emph{support}. While we now know that \emph{partial} support suffices, a key open question is whether the support can be removed \emph{entirely} while
retaining $O(d^{3/2})$ rounds for $d=n^\alpha$ with fixed $0<\alpha<2/3$?
Our algorithm uses public masks to choose clusters and balance communication
before execution; without support, these decisions must be made from
distributed information while respecting both send and receive limits.
An intermediate setting is \emph{triangle coverage}, where every potential
product $A_{ij}B_{jk}$ for a requested entry $X_{ik}$ has at least one public
coordinate among $(i,j)$, $(j,k)$, and $(i,k)$.

The new round complexity $O(d^{3/2})$ is certainly more elegant and natural than the prior upper bounds. It would be tempting to conjecture that $O(d^{3/2})$ is the right complexity for this problem, but there is no lower bound (even conditional lower bound). Hence a key open question is if the exponent $3/2$ can be reduced even with full support?
Improving this term is especially relevant for $d<n^{2/3}$: for larger $d$,
the known $O(dn^{1/3})$ algorithm already matches or improves our bound.

\begin{figure}[!ht]
	\centering
	\begingroup
	
	\def\omegaBound{9/4}
	\def\currentPower{1.5}
	\def\currentSwitchLabel{\frac23}
	\def\oldSemiringPower{1.867}
	\def\oldFieldPower{1.832}
	\pgfmathsetmacro{\currentSwitch}{1/(3*(\currentPower-1))}
	\pgfmathsetmacro{\oldSemiringSwitch}{1/(3*(\oldSemiringPower-1))}
	\pgfmathsetmacro{\oldFieldSwitch}{1/(3*(\oldFieldPower-1))}
	\pgfmathsetmacro{\densePower}{2-2/(\omegaBound)}
	\pgfmathsetmacro{\denseSwitch}{\densePower-1/3}
	
	\definecolor{boundblue}{HTML}{0088CC}
	\definecolor{boundorange}{HTML}{F26924}
	\definecolor{resultpurple}{HTML}{7744A3}
	\definecolor{conditionalorange}{HTML}{C65C16}
	\colorlet{unconditionalfill}{boundorange!34}
	\colorlet{conditionalfill}{boundorange!18}
	
	\newcommand{\DrawXTick}[3][0pt]{%
		\draw[thin] (#2,0) -- (#2,-.018)
		node[below=3pt,xshift=#1,font=\small] {$#3$};
	}
	\newcommand{\DrawBoundAxes}[1]{%
		\node[font=\normalsize] at (.5,1.77) {#1};
		\draw[axis] (0,0) -- (1.055,0);
		\draw[axis] (0,0) -- (0,1.56);
		\node[anchor=west,font=\small] at (0,1.63) {$T=n^y$};
		\node[anchor=north east,font=\small] at (1.055,-.22) {$d=n^x$};
		\DrawXTick{0}{0}
		\DrawXTick[-2pt]{\currentSwitch}{\currentSwitchLabel}
		\DrawXTick{1}{1}
		\foreach \y/\lab in {{1/3}/{\frac13},.5/{\frac12},1/1,{4/3}/{\frac43},1.5/{\frac32}}
		\draw[thin] (0,\y) -- (-.013,\y)
		node[left=2pt,font=\small] {$\lab$};
	}
	
	\resizebox{\linewidth}{!}{%
		\begin{tikzpicture}[x=6.2cm,y=4cm,
			axis/.style={thick,->},
			guide/.style={black!30,thin,densely dashed},
			ours/.style={resultpurple,line width=2.1pt,solid},
			prior sparse/.style={black!80,line width=.9pt,dashed},
			cube/.style={black!65,line width=.9pt,dashed},
			dense/.style={black!55,line width=.8pt,densely dashed},
			unconditional/.style={boundorange,line width=1pt},
			conditional/.style={conditionalorange,line width=1pt,dashed},
			curve label/.style={font=\normalsize,sloped},
			lower label/.style={font=\small,sloped,below=3pt}]
			
			\begin{scope}
				\coordinate (Origin) at (0,0);
				\coordinate (OldJoin) at (\oldSemiringSwitch,{\oldSemiringSwitch+1/3});
				\coordinate (NewJoin) at (\currentSwitch,{\currentSwitch+1/3});
				\coordinate (CubeStart) at (0,1/3);
				\coordinate (CubeEnd) at (1,4/3);
				
				\fill[boundblue!25]
				(Origin) -- (NewJoin) -- (CubeEnd) -- (1,1.5) -- (0,1.5) -- cycle;
				\fill[resultpurple!30] (Origin) -- (OldJoin) -- (NewJoin) -- cycle;
				\fill[unconditionalfill] (Origin) -- (1,1) -- (1,0) -- cycle;
				\fill[conditionalfill] (.4,.4) -- (CubeEnd) -- (1,1) -- cycle;
				\draw[guide] (\currentSwitch,0) -- (NewJoin);
				
				\draw[prior sparse] (Origin) --
				node[curve label,pos=.45,above=4pt,fill=boundblue!25,inner sep=1pt] {$d^{\oldSemiringPower}$}
				(OldJoin);
				\draw[cube] (CubeStart) --
				node[curve label,pos=.80,above=4pt,fill=boundblue!25,inner sep=1pt] {$dn^{1/3}$} (CubeEnd);
				\draw[ours] (Origin) --
				node[curve label,pos=.67,below=3pt,fill=white,inner sep=1pt] {$\boldsymbol{d^{\currentPower}}$}
				(NewJoin);
				\draw[unconditional] (Origin) --
				node[lower label,pos=.68,text=boundorange] {$\widetilde{\Omega}(d)$} (1,1);
				\draw[conditional] (.4,.4) --
				node[lower label,pos=.72,text=conditionalorange,inner sep=1pt]
				{$\widetilde{\Omega}(d^{14/9}/n^{2/9})$}
				(CubeEnd);
				
				\DrawBoundAxes{(a) Semirings}
				\DrawXTick{.4}{\frac25}
			\end{scope}
			
			\begin{scope}[xshift=8.4cm]
				\coordinate (Origin) at (0,0);
				\coordinate (OldJoin) at (\oldFieldSwitch,{\oldFieldSwitch+1/3});
				\coordinate (NewJoin) at (\currentSwitch,{\currentSwitch+1/3});
				\coordinate (DenseJoin) at (\denseSwitch,\densePower);
				\coordinate (RightEnd) at (1,\densePower);
				\coordinate (CubeStart) at (0,1/3);
				\coordinate (CubeEnd) at (1,4/3);
				
				\fill[boundblue!25]
				(Origin) -- (NewJoin) -- (DenseJoin) -- (RightEnd) --
				(1,1.5) -- (0,1.5) -- cycle;
				\fill[resultpurple!30] (Origin) -- (OldJoin) -- (NewJoin) -- cycle;
				\fill[unconditionalfill] (Origin) -- (1,1) -- (1,0) -- cycle;
				\draw[guide] (\currentSwitch,0) -- (NewJoin);
				\draw[guide] (\denseSwitch,0) -- (DenseJoin);
				\draw[guide] (0,\densePower) -- (RightEnd);
				
				\draw[prior sparse] (Origin) --
				node[curve label,pos=.45,above=4pt,fill=boundblue!25,inner sep=1pt] {$d^{\oldFieldPower}$}
				(OldJoin);
				\draw[cube] (CubeStart) --
				node[curve label,pos=.52,above=4pt,fill=boundblue!25,inner sep=1pt] {$dn^{1/3}$} (CubeEnd);
				\draw[ours] (Origin) --
				node[curve label,pos=.67,below=3pt,fill=white,inner sep=1pt] {$\boldsymbol{d^{\currentPower}}$}
				(NewJoin);
				\draw[dense] (DenseJoin) --
				node[font=\small,pos=.45,below=3pt,text=black!75,fill=white,inner sep=1pt]
				{$n^{2-2/\omega}$} (RightEnd);
				\draw[unconditional] (Origin) --
				node[lower label,pos=.68,text=boundorange] {$\widetilde{\Omega}(d)$} (1,1);
				
				\DrawBoundAxes{(b) Fields}
				\DrawXTick[2pt]{\denseSwitch}{\frac53-\frac2\omega}
				\draw[thin] (0,\densePower) -- (-.013,\densePower)
				node[left=2pt,yshift=2pt,font=\scriptsize] {$2-\frac2\omega$};
			\end{scope}
		\end{tikzpicture}%
	}
	\endgroup
	\caption{Purple region highlights our improvement. Dashed lines are prior algorithms, whose achievable region is blue.
		Orange lines are lower bound results from \cite{gupta-korhonen-etal-2025-low-bandwidth-matrix}. Results are summarized in \Cref{tab:us-full-support-bounds}. Logarithmic factors are suppressed. For field results we use $\omega\le9/4$.}
	\label{fig:us-full-support-bounds}
\end{figure}
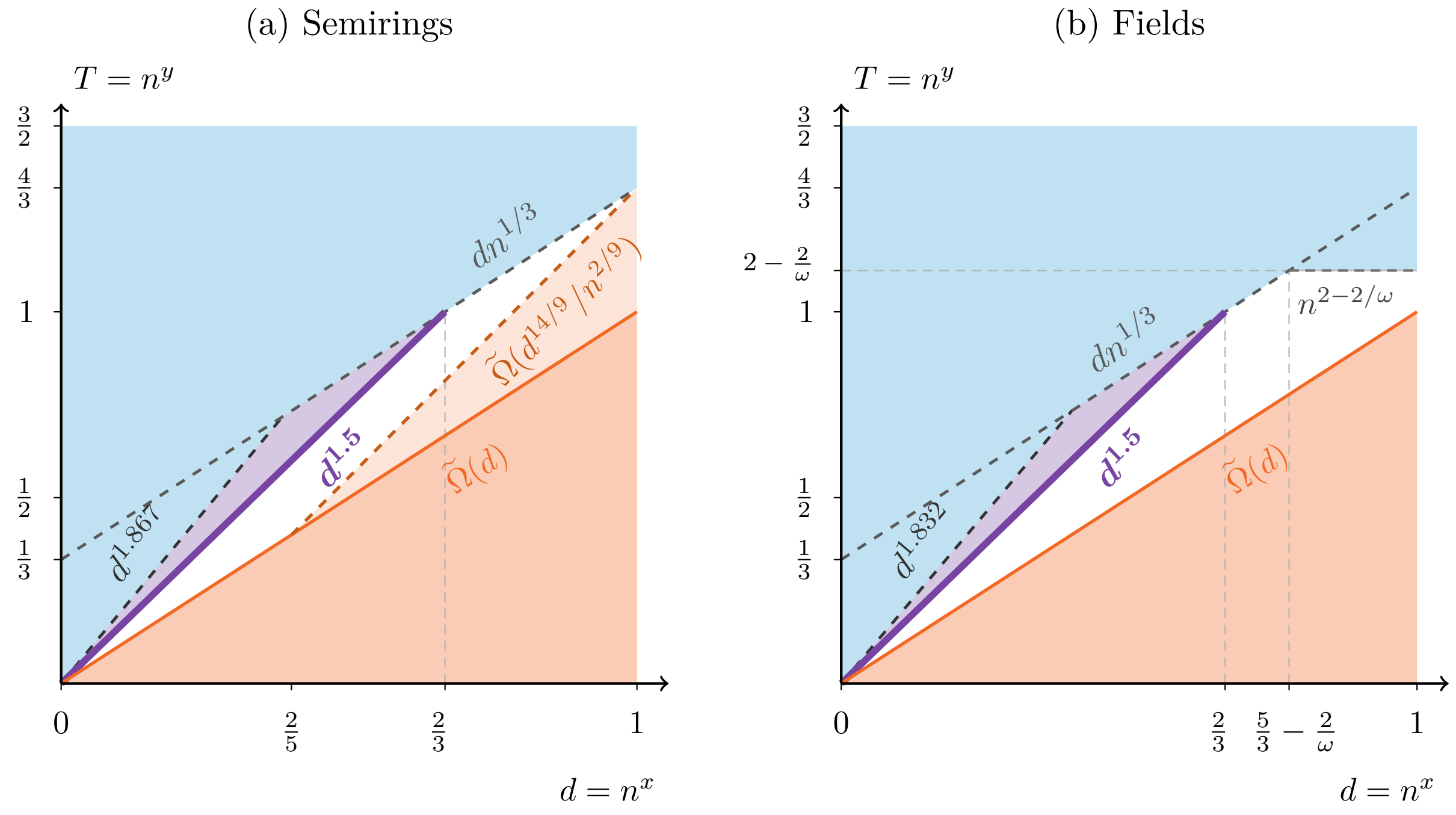

\paragraph{Overview of prior work and the new results.}
\Cref{tab:us-full-support-bounds,fig:us-full-support-bounds} summarize
the known bounds for uniformly sparse matrix multiplication with full
support and a public input distribution. Here $\omega$ denotes the
matrix multiplication exponent over the field in question. In the table, the dense matrix multiplication bound has been calculated using the latest result $\omega\le9/4$ \cite{openai-2026-an-upper-bound-of-9-4-for-the-matrix,amanov-van-den-berg-zuiddam-2026-multiplying-matrices-via-polynomials}, the corresponding bilinear algorithms~\cite{pass-2026-matrix-multiplication-from-polynomial-packings}, and the dense multiplication algorithm from \cite{censor-hillel-kaski-etal-2019-algebraic-methods-in-the}, resulting in the round complexity
$O(n^{10/9+\varepsilon})$ for every fixed $\varepsilon>0$.

\begin{table}[!ht]
	\centering
	\small
	\setlength{\tabcolsep}{4pt}
	\renewcommand{\arraystretch}{1.22}
	\begin{tabular}{@{}lcc@{}}
		\toprule
		Result & Semirings & Fields \\
		\midrule
		Trivial & $O(d^2)$ & $O(d^2)$ \\
		Dense multiplication~\cite{censor-hillel-kaski-etal-2019-algebraic-methods-in-the}
		& $O(n^{4/3})$ & $O(n^{10/9+\varepsilon})$ \\
		Sparse multiplication~\cite{censor-hillel-dory-etal-2021-fast-approximate-shortest}
		& $O(dn^{1/3})$ & $O(dn^{1/3})$ \\
		Gupta et al.\ (2022)~\cite{gupta-hirvonen-etal-2022-sparse-matrix-multiplication}
		& $O(d^{1.927})$ & $O(d^{1.907})$ \\
		Gupta et al.\ (2025)~\cite{gupta-korhonen-etal-2025-low-bandwidth-matrix}
		& $O(d^{1.867})$ & $O(d^{1.832})$ \\
		\textbf{This work} & $\boldsymbol{O(d^{1.5})}$ & $\boldsymbol{O(d^{1.5})}$ \\
		\midrule
		Best combined bound
		& $O(\min\{d^{1.5},dn^{1/3}\})$
		& $O(\min\{d^{1.5},dn^{1/3},n^{10/9+\varepsilon}\})$ \\
		\bottomrule
	\end{tabular}
	\caption{Summary of deterministic upper bounds for uniformly $d$-sparse matrix multiplication with support. The final row takes the best available bound in all regimes.}
	\label{tab:us-full-support-bounds}
\end{table}

\section{Preliminaries} \label{sec:prelim}
In this section we provide the definition of the model and problem, as well as a selection of helpful results and observations in the low-bandwidth model from prior works \cite{gupta-hirvonen-etal-2022-sparse-matrix-multiplication, gupta-korhonen-etal-2025-low-bandwidth-matrix, gupta-suomela-vahidi-2026-rectangular-matrix}.

\subsection{Model and Problem Definitions} \label{subsec:definitions}

We work in the \emph{low-bandwidth model}, a distributed system that consists of $n$ machines, denoted by
$M = \{1,2,\dots,n\}$, where all pairs of machines are directly connected.
Computation proceeds in synchronous communication rounds. And, for the purposes of this work, we only consider deterministic algorithms.

\subparagraph{Local computation.}
Each machine has unbounded local memory and local computation is free. The complexity measure of interest is the number of communication rounds.

\subparagraph{Communication.}
In each round, every machine may
\begin{itemize}
	\item send at most one $O(\log n)$-bit message to another machine, and
	\item receive at most one $O(\log n)$-bit message from another machine.
\end{itemize}
Equivalently, in each round the communication forms a directed graph of maximum in-degree and out-degree~$1$. An algorithm is invalid if $2$ messages are sent to a machine in any round.

\subparagraph{Problem.}
We study uniformly sparse matrix multiplication. Given two $n\times n$ matrices $A,B$, the task is to compute specified entries of their product $X=AB$.

The instance includes binary masks $\bar{A},\bar{B},\bar{X}\in\{0,1\}^{n\times n}$ that mark either allowed input entries or requested output entries; all other input entries are zero.
We may occasionally abuse notation and identify a mask with the set of its marked positions, and write
\[
  Q=\{(i,k)\in[n]\times[n]:\bar{X}_{ik}=1\}.
\]
A matrix or a mask is \emph{uniformly $d$-sparse} if each row and each column contains at most $d$ marked positions.
We denote the class of such masks by $\US(d)$ and assume that $\bar{A},\bar{B},\bar{X}\in\US(d)$, where $d\in[n]$.
Note that the full product $X$ need not be sparse; however, $\bar{X}$ is guaranteed to be.

Throughout the paper, matrix multiplication is performed over an arbitrary semiring $(\mathcal{S},+,\cdot)$. We assume that the input entries, products, and partial sums used by the algorithms admit encodings of size $O(\log n)$ bits.
In particular, each such value fits within one communication message in the low-bandwidth model.

\subparagraph{Input and output distribution.}
The input is evenly distributed among the $n$ machines, and each machine is responsible for producing an evenly sized portion of the output.
More concretely:
\begin{itemize}
  \item each machine initially holds at most $O(d)$ input entries in total;
  \item each machine is required to output at most $O(d)$ entries of $X$ requested by $\bar{X}$.
\end{itemize}
The algorithm may choose the balanced assignment of requested output positions to machines. 

\subparagraph{Supported setting.}
In the supported setting, the input masks, the output mask, and the assignment of input values to machines are fixed in advance and known to all machines.
The output assignment is chosen from this public information before the input values are revealed and is also known to all machines.
The numerical values at marked input positions are initially supplied to their holders and are not part of this public information.
All machines know $n$ and $d$.

\subparagraph{Complexity.}
An algorithm runs in $T(n,d)$ rounds if, after $T(n,d)$ communication
rounds, all machines have produced their assigned output.

\subsection{Observations and Toolbox}\label{subsec:toolbox}

For any integer $n>0$, let $[n]$ denote the set $\{1,\ldots,n\}$.

We next briefly mention some useful tools. Note that machines do not coordinate during the algorithm to determine which machine they need to send a message to, or which machine they need to receive one from. Instead, the scheduling is already hard-coded into the algorithm; during execution, the values are revealed and transferred. This avoids the costly runtime coordination overhead. 

\begin{lemma}[Broadcast~{\cite[Lemma 6.9]{gupta-korhonen-etal-2025-low-bandwidth-matrix}}]
	\label{lem:broadcast}
	In the low-bandwidth model, broadcasting a single bit $v \in \{0,1\}$ to all machines takes $\Theta(\log n)$ rounds. Moreover, if $k>1$ messages are given, broadcasting can be done in $O(k+ \log n)$ rounds.
\end{lemma}
\begin{proof}
	We can broadcast in a power-jumping manner. Suppose machine~$1$ initially holds the value $v$. Before round~$i$, the first $2^{i-1}$ machines know $v$. Each of them sends $v$ to a distinct machine that does not yet know it, until all machines have received it. For example, machine~$j$ sends to machine~$j+2^{i-1}$ whenever $j+2^{i-1}\leq n$. After round~$i$, $\min\{2^i,n\}$ machines know $v$, so all machines know it after $O(\log n)$ rounds.
	
	This argument works for any value $v$ that fits in a single message. By carefully interleaving broadcasts of this form~\cite{kannan-ballard-park-2016-a-high-performance-parallel}, we can schedule the messages in a pipeline and achieve $O(k+\log n)$ rounds.
	
	We refer the reader to \cite[Lemma 6.9]{gupta-korhonen-etal-2025-low-bandwidth-matrix} for details of the lower bound.
\end{proof}

\begin{lemma}[Aggregation~{\cite[Lemma 6.4]{gupta-korhonen-etal-2025-low-bandwidth-matrix}}]
	\label{lem:aggregation}
	In the low-bandwidth model where all $n$ machines are given a single bit in $\{0,1\}$, computing logical AND, OR, and sum of the values takes $\Theta(\log n)$ rounds. Moreover, if machines are instead given $k$ values that fit in a message, aggregation can be done in $O(k + \log n)$ rounds.
\end{lemma}
\begin{proof}
	The idea is the same as in \Cref{lem:broadcast}, with the communication done in reverse order. This also works for values that fit in a message, provided that the intermediate results fit in a message as well. The same pipelining idea of \cite{kannan-ballard-park-2016-a-high-performance-parallel} extends the argument to $k$ values.
	
	We refer the reader to \cite[Lemma 6.4]{gupta-korhonen-etal-2025-low-bandwidth-matrix} for details of the lower bound.
\end{proof}

We occasionally use the term \emph{distribution} $D$ of data to specify which machine $m\in M$ holds which value $v$, such as an element of a matrix. Typically, $v$ is small enough to fit in a single message, i.e., $O(\log n)$ bits.

\begin{lemma}[Redistribution \cite{gupta-suomela-vahidi-2026-rectangular-matrix}]\label{lem:redistribution-offline}
	Given globally known distributions $D$ and $D'$ of $d$ values per machine, we can rearrange the values from $D$ to $D'$ in $d$ rounds of the low-bandwidth model.
\end{lemma}
\begin{proof}
	Consider the bipartite multigraph of senders and receivers, where each value forms an edge between its home machine in $D$ and its destination machine in $D'$. This multigraph is $d$-regular and therefore admits a proper edge coloring with $d$ colors. We iterate over the color classes, sending all messages of one color class in each round. Since each color class is a matching, every machine sends and receives at most one message per round. As $D$ and $D'$ are globally known, all machines can compute the same coloring locally using a fixed deterministic procedure, without communication.
\end{proof}

\section{Fast Multiplication with Support} \label{sec:ub}
In this section we prove our main result, an $O(d^{3/2})$-round sparse matrix multiplication algorithm. We first give a bird's-eye view of the algorithm and then prove the following.

\begin{theorem}\label{thm:fs-upper-bound}
	There is a deterministic algorithm that solves any instance of uniformly sparse matrix multiplication in $O(\upbound)$ rounds in the supported low-bandwidth model.
\end{theorem}

\subsection{Overview}
Let $Q$ be the set of requested output positions in $\bar{X}$. We plan to partition $Q$ into two parts: (i) a set of clusters of many positions bounded in a small rectangle in $\bar{X}$, (ii) the remainder of unclustered positions that now have an extra sparsity structure. Then we process these positions separately and send the results or subproducts (i.e. contributions to a requested position of $X$) to the final holder of the output assigned in the distribution.

\paragraph{Step 1.~ Find clusters of densely packed positions.}
Set $t=d^{3/2}$. Repeatedly select $t$ requested
positions of $Q$ lying in at most $d$ rows and $d$ columns (i.e. a rectangle), form a cluster, and remove these positions from $Q$.
This produces at most $n/\sqrt d$ clusters. Let $Q_r$ be the set of remaining
positions, which contains fewer than $t$ positions
in every such rectangle.

We choose one public output assignment for both parts, balancing
output counts and the contributions received for $Q_r$.

\paragraph{Step 2.~Compute the clustered entries.}
Give each cluster a team of $\sqrt d$ machines and process all clusters in parallel. Assign the middle indices
$j$ to team members so that their input loads are balanced: a member that is assigned $j$ requires entries of column $j$ of $A$ and row $j$ of $B$, $2d$ in total. Moreover, the greedy assignment gives each member $O(d^{3/2})$ input entries.

\paragraph{Step 3.~Compute the remaining entries.}
Machine $j$ collects column $j$ of $A$ and row $j$ of $B$. Now we take advantage of the new sparsity structure: all possible subproducts that $j$ computes belong to a rectangle of $d$ rows and $d$ columns; therefore it has fewer than $t$ subproducts to compute. Then machine $j$ sends these products directly to their output holders, where the holders add the contributions from all middle indices. Each machine sends fewer than $t$ products, and our output assignment ensures
that each holder receives $O(t)$ products. Thus this phase
also takes $O(t)$ rounds.

\subsection{Proof of \Cref{thm:fs-upper-bound}}

\subsubsection{Step 1.~Clustering dense rectangles of $\bar{X}$.}
We set the threshold parameter $t$ to $d^{3/2}$. Then for every pair of subsets $I',K' \subseteq [n]$ of indices such that $|I'|,|K'| \leq d$, consider $Q\cap(I'\times K')$, i.e.~ a rectangular submatrix of $X$: check whether there are at least $t$ marked positions; if so, choose exactly $t$ of them, form a cluster $C_1$ with these $t$ positions and remove them from $Q$. Repeat this process until there are no such dense submatrices and form the set of clusters $\clusters =\{C_1,C_2,\ldots \}$. Let $Q_r$ be the remaining unclustered marked positions in $Q$. The key observation is that machines can compute these clusters in a globally consistent manner.

\begin{observation}[clustering without communication]
	The above clustering process can be done deterministically and individually by machines without communication; all machines will compute $\clusters$ and $Q_r$. The only information needed is the global support. Then every selection of indices $I'$, $K'$, and marked positions of $Q\cap(I'\times K')$ is done in lexicographical order. This ensures unique $\clusters$ and $Q_r$.
\end{observation}
	For the remainder of this section assume that all machines have computed $\clusters$ and $Q_r$. We can show that there are few clusters:
\begin{lemma}
	There are at most $n/\sqrt d$ clusters of size exactly $d^{3/2}$.
\end{lemma}
\begin{proof}
	The original output mask $\bar{X}$ contains at most $nd$ marked positions.
	Each cluster removes exactly $t$ distinct positions; therefore
	$|\clusters|\le nd/t$. Also, $|Q_r|\le nd$.

\end{proof}

\paragraph{Output assignment.}
For each $(i,k)\in Q_r$, use the number of indices $j$ with
$\bar A_{ij}=\bar B_{jk}=1$ as its weight. Order all requested positions of the original mask $\bar X$ by decreasing weight, breaking ties lexicographically, and assign them cyclically to machines $1,\ldots,n$.
Each machine gets at most $|\bar X|/n=nd/n\le d$ positions.
This assignment is computed locally without communication and
kept fixed throughout the algorithm.

\subsubsection{Step 2.~Compute the clustered entries.}
Give each cluster $C\in\clusters$ a distinct and disjoint team
$T_C=\{m_1,\ldots,m_{\sqrt d}\}$ of helper machines. We can assign them deterministically by their id: let the first $\sqrt{d}$ machines be the helper team for the first cluster, and so on.
Let $I_C$ and $K_C$ be the row and column indices occurring in
the selected positions of $C$. Note that both sets have size at most $d$.
We assign each middle index $j\in[n]$ to one team member and
write $J(C,m_p)$ for the indices assigned to $m_p$.
Each member will compute partial sums for positions $(i,k)\in C$
using its assigned middle indices. Assigning middle index $j$ to a member means assigning it the marked entries $A_{ij}$ with $i\in I_C$ and
$B_{jk}$ with $k\in K_C$. The following lemma shows that we can spread the work between machines without congesting a node.

\begin{lemma}[greedy assignment]\label{lem:greedy-assignment}
	The sets $J(C,m_p)$ can be chosen deterministically and locally,
	without communication, so that they partition $[n]$ and each
	team member requires at most $O(d^{3/2})$ marked input entries.
\end{lemma}
\begin{proof}
	Fix a cluster $C$. Denote by
	\[
	  c_j=
	  |\{i\in I_C:\bar A_{ij}=1\}|
	  +|\{k\in K_C:\bar B_{jk}=1\}|,
	\]
	the number of marked entries that $j$ participates in. Each $c_j\le 2d$. Moreover, the marked entries in rows $I_C$
	of $A$ and columns $K_C$ of $B$ number at most $2d^2$ in total,
	so $\sum_j c_j\le 2d^2$.

	We assign the middle indices in increasing order. Assign each
	index to a team member with the fewest currently assigned
	input entries. Before an
	assignment, this minimum load is at most the average load
	and hence at most $2d^2/\sqrt d$. The assignment adds at most
	$2d$ entries. Thus every team member requires at most
	\[
	  \frac{2d^2}{\sqrt d}+2d=2d^{3/2}+2d = O(d^{3/2})
	\]
	entries. All counts and choices depend only on
	the public masks, so every machine can compute the assignment
	locally without communication.
\end{proof}

\begin{lemma}\label{lem:compute-clustered}
	All output positions in the clusters $\clusters$ can be
	computed and delivered to their assigned output holders in
	$O(\upbound)$ rounds.
\end{lemma}
\begin{proof}
	By \Cref{lem:greedy-assignment}, each team member needs at most
	$L=O(d^{3/2})$ input entries.

	\paragraph{Gathering the inputs.}
	An entry $A_{ij}$ is needed only by clusters containing row $i$,
	and by one member per cluster. Since the clusters are disjoint
	and row $i$ has at most $d$ requested positions, there are at most
	$d$ recipients. The same argument applies to $B_{jk}$ using
	column $k$. Let $f_e\le d$ be the number of recipients of each
	needed entry $e$, and let $R=\sum_e f_e\leq nL$.

	Process these entries in a fixed publicly known order. Assign each
	to the broadcasting machine with the fewest currently assigned
	copies to send.
	Each assignment adds at most $d$ copies, so the same greedy
	argument gives at most $R/n+d\le L+d$ copies per broadcasting
	machine. It also needs at most $L+d$ distinct entries, since
	each assigned entry has at least one recipient.

	First, original holders send one copy of each needed entry to
	its broadcasting machine: each holder sends $O(d)$ values and
	each broadcaster receives at most $L+d$. The broadcasters then
	send the copies to the team members: each sends at most $L+d$
	copies and each member receives at most $L$. These schedules
	are public, so \Cref{lem:redistribution-offline} implements both
	stages in $O(L+d)=O(d^{3/2})$ rounds.

	\paragraph{Computing and delivering partial sums.}
	Each member $m_p\in T_C$ computes
	\[
	  S_{C,p}(i,k)=\sum_{j\in J(C,m_p)}A_{ij}B_{jk},
	  \qquad (i,k)\in C,
	\]
	and sends these values to the assigned output holders.
	Each member sends $|C|=d^{3/2}$ partial sums. Each holder has
	$O(d)$ output positions and receives $\sqrt d$ partial sums
	per clustered position, hence $O(d^{3/2})$ values in total.
	This redistribution also takes $O(d^{3/2})$ rounds. All clusters are
	processed in parallel, which gives $O(\upbound)$ rounds overall.
\end{proof}

\subsubsection{Step 3.~Compute the remaining entries.}
We compute the remaining positions of $Q_r$ by computing their intermediate products first, then adding those products at the final output holders.

A key observation here is that $Q_r$ now has additional sparsity structure: small rectangles in $Q_r$ have few unclustered positions.

\begin{observation}\label{obs:remainder-sparsity}
	At termination,
	\[
	|Q_r\cap(I'\times K')|<t
	\]
	for all $I',K'\subseteq[n]$ with $|I'|,|K'|\le d$.
	This follows directly from the stopping condition of the clustering.
\end{observation}

\begin{lemma}[output assignment]\label{lem:output-assignment}
	Each output holder receives fewer than $t+d$ contributions
	in Step~3.
\end{lemma}
\begin{proof}
	Let $L_r$ be machine $r$'s total assigned weight and let $P=\sum_r L_r$.
	Observation~\ref{obs:remainder-sparsity} gives fewer than $t$
	contributions for a fixed middle index, so $P<nt$. Each weight is
	at most $d$.

	The sorted cyclic order gives $L_1\ge L_r\ge L_n$,
	treating missing weights in the last cycle as zero.
	Every weight on machine $1$ after its first is bounded by
	the preceding weight on machine $n$. Hence
	\[
	  L_r\le L_1\le L_n+d\le P/n+d<t+d.
	\]
\end{proof}

\begin{lemma}\label{lem:compute-remainder}
	We can compute output positions $Q_r$ and route them to their assigned output holder in $O(\upbound)$ rounds.
\end{lemma}
\begin{proof}
	First, each original holder sends a copy of every marked entry
	$A_{ij}$ and $B_{jk}$ to machine $j$. Each holder sends $O(d)$
	values and each receiver gets at most $2d$. This takes $O(d)$
	rounds by \Cref{lem:redistribution-offline}.

	Machine $j$ computes $A_{ij}B_{jk}$ for each $(i,k)\in Q_r$
	that has both its input positions marked. There are at most $d$
	possible rows $i$ and $d$ possible columns $k$, so
	Observation~\ref{obs:remainder-sparsity} gives fewer than $t$
	such products.

	Send each product, tagged by $(i,k)$, directly to its assigned
	output holder.
	Each machine sends fewer than $t$ products, and by
	\Cref{lem:output-assignment} each holder receives fewer than
	$t+d$. This public redistribution takes $O(t+d)$ rounds. Then each holder locally adds the contributions for each of its positions
	in $Q_r$. The total cost is $O(d)+O(t+d)=O(\upbound)$ rounds.
\end{proof}

\begin{proof}[Proof of \Cref{thm:fs-upper-bound}]
	It follows by combining all three steps, and in particular \Cref{lem:compute-clustered} and \Cref{lem:compute-remainder}.
\end{proof}

\section*{AI Methodology}
We used OpenAI Codex, including GPT-6 Astra and GPT-6.1 Sol, during the research and preparation of this manuscript.
The tools assisted with algorithm design, its proof, identifying relevant
literature, and revising text. Moreover, we used AI to formalize the results of this paper in Lean.\footnote{Lean formalization is publicly available at \url{https://doi.org/10.5281/zenodo.24127514}}
The authors take full responsibility for the correctness and presentation of the results and for the accuracy of all references.

\bibliographystyle{plainurl}
\bibliography{da}

\end{document}